\documentclass[11pt]{article}
\usepackage[a4paper,margin=2.2cm]{geometry}
\usepackage[T1]{fontenc}
\usepackage[utf8]{inputenc}
\usepackage{mathptmx}
\usepackage{amsmath,amssymb,amsthm}
\usepackage{graphicx,booktabs,array,calc,tabularx,xurl}
\usepackage[dvipsnames]{xcolor}
\usepackage[most]{tcolorbox}
\usepackage[colorlinks=true,linkcolor=black,citecolor=black,urlcolor=blue!50!black,pdftitle={An Executable Audit of Software Reliability Growth Models: Validity, Fitted Shape, and Limits of Inference},pdfauthor={Quyet-Thang Huynh},pdfkeywords={software reliability growth models; executable audit; correctness; mean shape; statistical inference; reproducibility}]{hyperref}
\providecommand{\tightlist}{\setlength{\itemsep}{0pt}}
\providecommand{\real}[1]{#1}
\newtheorem{proposition}{Proposition}
\newtheorem{theorem}{Theorem}
\newenvironment{boxwithhead}[2]{\begin{tcolorbox}[breakable,colback=gray!4,colframe=gray!60,title={#1},fonttitle=\bfseries\small,fontupper=\small]#2}{\end{tcolorbox}}
\graphicspath{{./}}
\begin{document}
\title{An Executable Audit of Software Reliability Growth Models:\\Validity, Fitted Shape, and Limits of Inference}
\author{Quyet-Thang Huynh\\\small School of Information and Communication Technology, Hanoi University of Science and Technology, Ha Noi, Vietnam\\\small \texttt{thang.huynhquyet@hust.edu.vn}}
\date{}
\maketitle
\begin{abstract}
Applying a software reliability growth model (SRGM) requires three separate justifications: mathematical validity, the shape of the fitted mean, and claims about population shape. We present an executable audit for three existing SRGM rate families on explicit finite horizons. A declared contract specifies the assumptions, and the audit links each claim to its supporting evidence. Two routes serve different purposes. The record-validation route checks semantic and provenance obligations against trusted context. The certificate route separates bounded witness discovery from a checker that derives inflection counts and orientations without an expected-count input. The mathematical basis comprises analytical classification, conditional certificate soundness, and stability of separated simple roots; implementation trust boundaries are explicit. In a 180-request synthetic campaign, the certificate route certifies 169 of 174 admissible inputs and abstains on five. Every certified count and orientation agrees with an independently implemented reference. On a frozen mutation corpus, the audit rejects all 840 reference-labelled mathematical-disagreement records; four computational baselines combined reject 646. The audit nevertheless retains 424 changed-field integrity misses. We also calibrate one band/projection procedure on 1,000 simulated histories. Among 200 delayed-boundary histories, it returns 22 singleton count outputs, of which 11 are incorrect. Thus, correctly classifying a selected fitted curve does not ensure informative or calibrated population inference. Three published failure histories illustrate the reporting consequences. The audit makes validity and fitted-shape claims inspectable and replayable, while leaving unsupported population conclusions unresolved.\end{abstract}
\hypertarget{introduction}{%
\section{Introduction}\label{introduction}}

A logistic fit to the IBM Entry failure history shows why a software
reliability growth model (SRGM) needs more than a shape label. On the
14-day training horizon, the selected mean is convex and has no interior
inflection. Extending the same fitted curve to 21 days yields an
inflection at approximately day 14.554. Both classifications are
mathematically correct on their respective horizons, and the model
remains admissible. The change arises from the analysis horizon rather
than a new fit. Yet an inflection beyond the training horizon does not
establish that the observations support an inflection in the population
mean. This example, examined in Section 4, motivates a distinction
between what the model permits, what the fitted curve exhibits and what
the data establish.

These are three claim layers. \textbf{Mathematical validity} concerns
whether a declared construction satisfies its analytical conditions,
including rate nonnegativity over the stated domain. \textbf{Fitted-mean
shape} concerns the curvature and interior inflections of the selected
mean on a specified horizon. \textbf{Population-shape evidence} concerns
whether an inferential procedure supports a shape claim about the
underlying mean. Evidence at one layer does not establish a conclusion
at the next: an admissible model can have several possible shapes, and a
correctly classified fitted curve can have uncertain population shape.
SRGMs describe accumulated failures during testing or operation; under
an NHPP observation model, mean increments determine expected interval
counts and the mean derivative determines intensity. Classical models
\cite{ref1,ref2,ref3} and studies of assumptions and prediction \cite{ref4,ref5} provide
this context, but a reported shape still requires an explicit target
function, parameter domain and horizon.

The missing element is an executable link between the conditions for a claim and the decision reported for an application. Numerical convergence does not establish rate admissibility. Counting curvature zeros does not distinguish interior inflections from boundary or noncrossing zeros. Changing the horizon can also change fitted shape without adding statistical evidence. Analytical classification, numerical checking and uncertainty assessment therefore address different obligations. An audit must connect these obligations and retain an unresolved outcome when evidence is missing. Our audit object is \textbf{a declared SRGM application and its associated decision records}. Published histories motivate the reporting problem. Controlled mutations test the stated obligations, but do not estimate how often such errors occur in practice.

The central claim is that \textbf{validity and fitted-mean shape can be
audited under a declared model contract, while population-shape claims
require separate inferential evidence}. This claim determines both the
audit scope and the interpretation of its results.

We therefore ask three research questions:

\begin{itemize}
\tightlist
\item
  \textbf{RQ 1 --- Specification and guarantees:} Which obligations
  support mathematical validity and fitted-mean shape on a declared
  domain and horizon, and when must the audit abstain?
\item
  \textbf{RQ 2 --- Correctness and added value:} How reliably does the certificate route verify supported fitted-shape assertions, and what additional record-checking capability does the record-validation route provide over the computational baselines?
\item
  \textbf{RQ 3 --- Limits of the evaluated inference procedure:} When fitted shape is correctly
  determined, where does the working band/projection procedure fail to support a population-shape claim?
\end{itemize}

The paper makes three contributions.

\textbf{C1 --- A contract-based executable audit.} The audit checks three existing SRGM rate families under declared contracts. It links mathematical validity, horizon-specific fitted shape and separate population evidence to explicit obligations. The certificate route derives counts and orientations from checked witnesses without an expected-count input. The record-validation route checks semantic and provenance obligations against trusted context. Each decision retains its assumptions, evidence and unresolved obligations (Section 4 and Supplement S7).

\textbf{C2 --- Conditional correctness guarantees.} Analytical classification and certificate soundness connect the supported rate formulas to fitted shape on a finite horizon. A stability result covers finitely many separated simple roots. A conditional completeness result applies to the specified reference search. Analytical proofs and scoped Lean developments support these guarantees, subject to the premises and implementation boundaries in Section 3.

\textbf{C3 --- Evaluation of checking and inference limits.} We measure certificate correctness, yield, abstention and cost in three campaigns: 180 primary requests, 90 extension requests and 30 printed-point contracts. Frozen-corpus comparisons, ablation and benign controls assess record rejection and residual misses. A separate post-freeze challenge tests composite mutations. Calibration of one working inference procedure establishes its limits in the tested design; three published histories illustrate the reporting consequences (Sections 5--6).

\hypertarget{problem-definition-scope-and-audit-contract}{%
\section{Problem definition, scope and audit
contract}\label{problem-definition-scope-and-audit-contract}}

\hypertarget{modeling-setting}{%
\subsection{Modeling setting}\label{modeling-setting}}

For \(0\le t_0<T<\infty\), the declared construction is

\[m'=b(A-pm),\qquad m(t_0)=0,\]

where the declared contract supplies fixed positive \(A\) and \(p\), and \(b\) is a deterministic rate from a supported family. We check mathematical validity over the closed horizon \(H=[t_0,T]\) and count inflections only in its open interior. Empirical fitting fixes \(p=1\) and uses the frozen bounded numerical policy. For fitted-curve auditing, estimated parameters are treated as fixed inputs. Their statistical uncertainty requires separate inference. These mathematical checks do not establish that the observed increments satisfy the NHPP assumptions.

\begin{table}[!t]
\caption{Declared scope of the audit.}\label{tab:1}
\centering
\small
\begin{tabular}{@{}
  >{\raggedright\arraybackslash}p{(\textwidth - 4\tabcolsep) * \real{0.3333}}
  >{\raggedright\arraybackslash}p{(\textwidth - 4\tabcolsep) * \real{0.3333}}
  >{\raggedright\arraybackslash}p{(\textwidth - 4\tabcolsep) * \real{0.3333}}@{}}
\toprule\noalign{}
\begin{minipage}[b]{\linewidth}\raggedright
Component
\end{minipage} & \begin{minipage}[b]{\linewidth}\raggedright
Supported application
\end{minipage} & \begin{minipage}[b]{\linewidth}\raggedright
Outside this study
\end{minipage} \\
\midrule\noalign{}

Rate & Delayed-rational, logistic, rational-logistic & Arbitrary
families or uncertain parameter boxes \\
Dynamics & Constant positive \(A,p\), declared/reset origin &
Time-varying fault totals or uncertain \(p\) \\
Shape & Mean curvature on explicit finite horizon & Shape inferred only
from model name/rate \\
Fitting & Locked bounded policy, empirical \(p=1\) & Universal
global-optimality guarantee \\
Inference & Evaluation of the declared working procedure & Generally
calibrated confidence for all cases \\
Validation & Scoped record predicates and evidence & Every source-code
or empirical-study defect \\
\bottomrule
\end{tabular}
\end{table}

The logistic/rational-logistic source subclass uses \(c=r\)~\cite{ref8}.
Independent \(c,r\) are analytical coordinates and a sensitivity class,
not a claimed new model. Setting \(b=cf_\psi\) makes the mean depend on
\(A/p\) and \(pc\); fixing \(p\) avoids claiming independent
identification of all three parameters. A restart at \(t_0\) resets the
mean and accumulated-rate integral. It does not retrospectively validate
an earlier negative-rate segment.

\hypertarget{inputs-outputs-and-two-status-axes}{%
\subsection{Inputs, outputs and two status
axes}\label{inputs-outputs-and-two-status-axes}}

A contract is
\(\mathcal A=(\mathcal F,\theta,A,p,t_0,T,\mathcal O,\mathcal P)\),
identifying the rate family/parameters, origin/units, observation
assumptions and fitting/inference policy. Source restrictions, bound
changes and extended horizons are separately recorded sensitivities.
Associated records retain validity \(V\), shape \(S\), mean
representation \(R\) and component evidence \(E\), with witnesses,
diagnostics and source/code/result locators.

The report separates three claim layers. L1 states mathematical validity under the declared premises and records observational adequacy separately. L2 describes the selected mean on the stated horizon: its shape class, inflection count and crossing orientations. It also records endpoint and noncrossing zeros, or an undetermined result. L3 assesses evidence for population shape and remains unresolved when calibration or selection premises fail. Labels such as verified or numerically supported apply to specific components. Combining them does not produce a statistical confidence level.

Record status is reported separately from these claim layers. A record may pass its supported obligations, be accepted as a scoped unresolved report, be malformed, fail a required check, or receive an abstention. These outcomes describe the record. An encoding error or numerical failure does not, by itself, make the model invalid. Supplement S1/S3 gives the full status vocabulary and API details.

\hypertarget{obligations-and-error-taxonomy}{%
\subsection{Obligations and error
taxonomy}\label{obligations-and-error-taxonomy}}

The audit groups obligations by the conclusion they support.
\textbf{Domain/origin obligations} establish supported parameters, a
closed validity horizon and the correct restart.
\textbf{Curvature/horizon obligations} establish the mean-specific
contrast, complete zero filtering, sign changes and orientation.
\textbf{Representation/numerical obligations} establish requested
routes, feasibility and numerical evidence without converting solver
failure into model invalidity. \textbf{Provenance/evidence obligations}
bind the record to its trusted input and prevent stronger aggregate
labels than applicable evidence supports. \textbf{Population
obligations} identify the target, failure denominator, calibration and
selection conditions.

The mutation taxonomy follows these obligation groups. Counting endpoint or double zeros, or substituting rate shape for mean shape, violates curvature obligations. Forgetting a shifted origin violates domain obligations. Treating solver failure as model invalidity violates representation semantics. Corrupt locators and overstated evidence labels violate provenance or evidence obligations. Population claims based on fitted counts or singleton count sets require the population obligations. Random integrity mutations additionally test how the obligations are encoded. A changed field is not necessarily a mathematical error.

\hypertarget{mathematical-basis-and-conditional-guarantees}{%
\section{Mathematical basis and conditional
guarantees}\label{mathematical-basis-and-conditional-guarantees}}

\hypertarget{admissibility-and-curvature-transfer}{%
\subsection{Admissibility and curvature
transfer}\label{admissibility-and-curvature-transfer}}

For \(b\in C^1(H)\), let \(B_{t_0}(t)=\int_{t_0}^t b(s)\,ds\). The ODE
solution and derivatives are

\[\begin{aligned}
m(t)&=\frac Ap(1-e^{-pB_{t_0}(t)}),&m'(t)&=Ab(t)e^{-pB_{t_0}(t)},\\
m''(t)&=Ae^{-pB_{t_0}(t)}q_p(t),&q_p(t)&=b'(t)-pb(t)^2.
\end{aligned}\]

\begin{proposition}
Under constant \(A,p>0\) and \(m(t_0)=0\), this
solution is finite on \(H\) and is an NHPP-compatible nondecreasing mean
exactly when \(b\ge 0\) throughout \(H\). Mean curvature has the signs
and zeros of \(q_p\).
\end{proposition}

\begin{proof}
Continuity on compact \(H\) makes accumulated rate finite.
The multiplier \(Ae^{-pB_{t_0}}\) is strictly positive. Thus \(m'\ge 0\)
exactly when \(b\ge 0\), and curvature signs/zeros agree with \(q_p\). A
nonnegative locally finite intensity constructs an NHPP mean; it does
not verify the observation model. \end{proof}

\hypertarget{finite-horizon-shape}{%
\subsection{Finite horizon shape}\label{finite-horizon-shape}}

An interior inflection is an isolated curvature zero strictly inside \((t_0,T)\), with opposite signs on its two sides. An endpoint zero has no two-sided neighborhood within the horizon. A noncrossing zero, including a double zero, does not change the curvature sign. Count and orientation must also be distinguished. A single positive-to-negative crossing supports the orientation-qualified delayed S-shape; a negative-to-positive crossing does not. A mean with no interior roots may remain convex throughout the horizon, so a zero count does not imply concavity. An affine mean requires an identically zero contrast; numerical smallness is insufficient.

\begin{table}[!t]
\caption{Supported rates and mean-curvature decision rules.}\label{tab:2}
\centering
Write \(x=\beta e^{-rt}\) with positive \(c,r,\beta,p\).
\small
\begin{tabular}{@{}
  >{\raggedright\arraybackslash}p{(\textwidth - 4\tabcolsep) * \real{0.3333}}
  >{\raggedright\arraybackslash}p{(\textwidth - 4\tabcolsep) * \real{0.3333}}
  >{\raggedright\arraybackslash}p{(\textwidth - 4\tabcolsep) * \real{0.3333}}@{}}
\toprule\noalign{}
\begin{minipage}[b]{\linewidth}\raggedright
Family and rate
\end{minipage} & \begin{minipage}[b]{\linewidth}\raggedright
Contrast numerator
\end{minipage} & \begin{minipage}[b]{\linewidth}\raggedright
Interior decision
\end{minipage} \\
\midrule\noalign{}

Delayed: \(r^2t/(1+rt)\) & \(1-pr^2t^2\) & Filter \(t=1/(r\sqrt p)\) to
\((t_0,T)\) \\
Logistic: \(c/(1+x)\) & \(rx-pc\) & Filter
\(t=r^{-1}\log(r\beta/(pc))\) \\
Rational: \(c(1+kx)/(1+x)\) & \(P(x)\) below & Retain simple roots in
\((x(T),x(t_0))\) \\
\bottomrule
\end{tabular}
\end{table}

For rational-logistic, the domain predicate and numerator are

\[k\ge-e^{rt_0}/\beta,\qquad
P(x)=r(1-k)x-pc(1+kx)^2,\]

and \(D=r(1-k)[r(1-k)-4pck]\). For \(k<0\), the numerator of the rate is
smallest at \(t_0\); for \(k\ge 0\) it is positive. Equality allows a
left-endpoint rate zero. With \(k_c=r/(r+4pc)\), \(k=0\) is linear;
admissible negative \(k\) allows at most one crossing; \(0<k<k_c\) can
allow two; \(k=k_c\) has a possible noncrossing double zero; larger
\(k\) has negative curvature. Real discriminant alone does not imply
positive roots or horizon membership. The complete branch table, stable
root expressions and primitives are in Supplement S2.

\begin{proposition}
Under Proposition 1, complete isolation and
correct neighboring signs of the supported numerator's zeros determine
the fitted inflection count by retaining precisely the sign-changing
zeros in \((t_0,T)\). For rational-logistic these are simple positive
roots satisfying \(\beta e^{-rT}<x_i<\beta e^{-rt_0}\). The count is at
most two; endpoints and double zeros contribute zero.
\end{proposition}

\begin{proof}
Positive multipliers/denominators preserve signs.
Continuity fixes the sign between consecutive isolated zeros. Simple
linear/quadratic roots cross sign, whereas double roots do not. The
decreasing map \(x=\beta e^{-rt}\) gives the strict horizon
inequalities. The affine case requires separate treatment. \end{proof}

Mean and rate shape can differ: \(b=t/(1+t)\) has no rate inflection but
its mean changes curvature at \(t=1\) for \(p=1\). Rational-logistic
\(c=r=p=1,\beta=100,k=0.1\) has increasing rate and two mean
inflections, approximately 0.377738 and 4.227432. These are analytical
counterexamples to an implication, not assertions that prior authors
made that implication.

\hypertarget{certificate-soundness-and-abstention}{%
\subsection{Certificate soundness and
abstention}\label{certificate-soundness-and-abstention}}

\begin{theorem}
Let \(q\in C^1(H)\). Suppose finitely many pairwise
disjoint interior root boxes \(J_i=[a_i,b_i]\) have strictly opposite
endpoint signs and derivatives of constant strict sign throughout each
box. Suppose every other interior point is covered by intervals on which
\(q\) is proved nonzero. Then each box contains exactly one
sign-changing zero and no further interior zero exists.
\end{theorem}

\begin{proof}
The intermediate value theorem gives a zero in each box.
The derivative premise and mean value theorem give strict monotonicity
and uniqueness, with opposite signs on the two sides. Disjointness
separates the roots, and complete zero-free complement coverage excludes
others, yielding a bijection from boxes to interior zeros. \end{proof}

Valid enclosures over whole intervals can establish the required sign and derivative bounds. A grid alone cannot exclude zeros from the complement of the root boxes. An exactly proved global-endpoint zero contributes no interior zero when the required one-sided monotonicity holds. Double roots need a separate exact factor or multiplicity witness. The checker abstains if a comparison is unproved or coverage is incomplete. The theorem establishes conditional soundness; it does not automatically construct a certificate for every rate.

\hypertarget{trust-boundary}{%
\subsection{Trust boundary}\label{trust-boundary}}

Lean 4.8.0 with pinned Mathlib checks real-valued certificate
cardinality, crossings, endpoint exclusion and supporting factor lemmas.
It also derives \(m''=(A-pm)(b'-pb^2)\) from differentiability and the
ODE, using the explicit premise \(A-pm>0\) for curvature transfer. It
does not prove the NHPP integral solution or global residual positivity.
The final module has 25 named supporting theorems; its accepted log and
two rejected negative assertions are preserved in Supplement S2.

\begin{table}[!t]
\caption{Evidence targets and trust boundaries.}\label{tab:3}
\centering
\small
\begin{tabular}{@{}
  >{\raggedright\arraybackslash}p{(\textwidth - 4\tabcolsep) * \real{0.3333}}
  >{\raggedright\arraybackslash}p{(\textwidth - 4\tabcolsep) * \real{0.3333}}
  >{\raggedright\arraybackslash}p{(\textwidth - 4\tabcolsep) * \real{0.3333}}@{}}
\toprule\noalign{}
\begin{minipage}[b]{\linewidth}\raggedright
Evidence
\end{minipage} & \begin{minipage}[b]{\linewidth}\raggedright
Supports
\end{minipage} & \begin{minipage}[b]{\linewidth}\raggedright
Additional premise or limit
\end{minipage} \\
\midrule\noalign{}

Analytical derivation & Mathematical transitions under assumptions & No
proof of every implementation \\
Reference/interval checks & Specific predicates in supported inputs &
Exact algebra/libraries and encoding scope \\
Lean theorem & Certificate soundness from supplied premises & No
automatic proof of every Python output \\
Calibration & Inference behavior in tested design & No universal or
selective-confidence guarantee \\
\bottomrule
\end{tabular}
\end{table}

The formalization establishes real-valued certificate soundness under supplied premises. The concrete parser, enclosure backend and Python checker remain outside its scope, and no case-specific Python-to-Lean bridge is provided. Statistical calibration is assessed separately in E3. Supplement S2.4 and S8.3 identify the formal targets and supporting evidence.

\subsection{Stability of separated simple roots and conditional reference-search completeness}
\label{multi-root-reference}
\begin{theorem}
Let $q,\widetilde q\in C^1(H)$ and let finitely many strictly separated interior boxes $J_i=[a_i,b_i]$ have opposite endpoint signs and $|q'|\ge\nu_i>0$ on each box. Put $K=H\setminus\bigcup_i(a_i,b_i)$ and choose margins satisfying
\[
 \begin{gathered}
 0<\eta_i\le\min\{|q(a_i)|,|q(b_i)|\}\quad(\text{each }i),\\
 |q(t)|\ge\gamma>0\quad(t\in K).
\end{gathered}
\]
If
\[
 \|\widetilde q-q\|_{\infty,H}\le\varepsilon<\gamma,\qquad
 \varepsilon<\eta_i,\qquad
 \|\widetilde q'-q'\|_{\infty,J_i}<\nu_i,
\]
then count, box membership, order and both crossing orientations are preserved, and $|\widetilde\tau_i-\tau_i|\le\varepsilon/\nu_i$. Opposite endpoint signs give existence; strict derivative margins give uniqueness and orientation; complete complement exclusion gives exhaustiveness. MVT at the two roots gives the displacement bound. With no root boxes, $\varepsilon<\gamma$ alone suffices. Perturbed rate admissibility is separate. Endpoint, tangent and zero-slope roots are outside this regular theorem; function sup-norm alone does not preserve uniqueness. Supplement S8.1 supplies the full statement, proof and rate-norm transfer.

\end{theorem}

The reference search eventually finds a checked regular cover under the following conditions: there are finitely many simple interior roots and no endpoint roots; the enclosures are sound and jointly locally convergent; and every effective call finishes. The search fairly enumerates dyadic windows, complement checks and precision levels. Overlapping windows accommodate roots that lie exactly on mesh splits. This is a conditional completeness result for the reference search. The archived Lean development checks the specified abstract interfaces, but no refinement theorem transfers eventual success to the bounded production generator. No uniform runtime bound is asserted. Supplement S8.2--S8.3 states the backend and encoding assumptions. These arguments specialize classical root-isolation and stability principles.

\hypertarget{audit-design-and-implementation}{%
\section{Audit design and
implementation}\label{audit-design-and-implementation}}

\hypertarget{workflow}{%
\subsection{Workflow}\label{workflow}}

\begin{figure}[!t]
\centering
\includegraphics[width=0.85\textwidth,keepaspectratio]{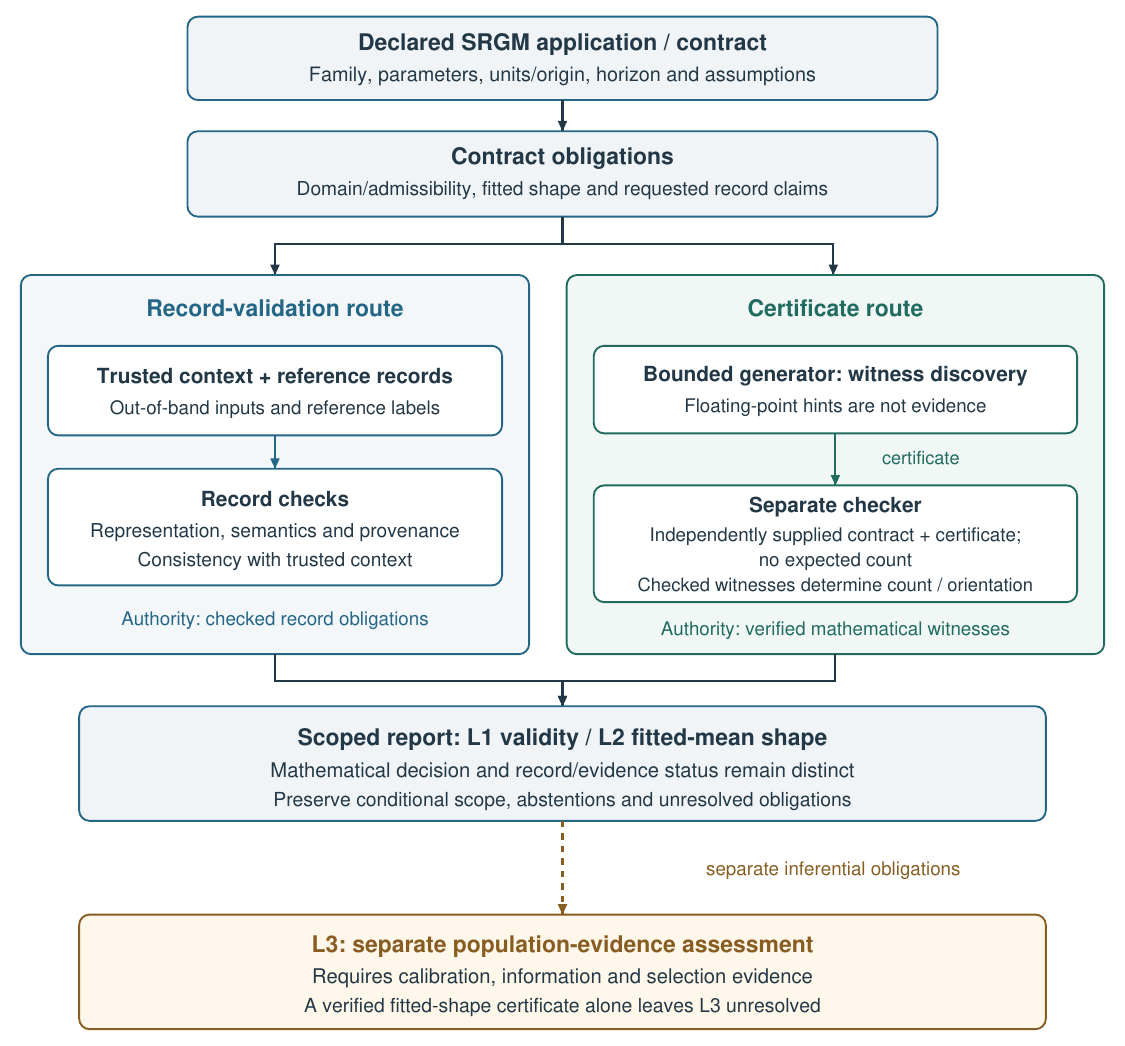}
\caption{Two execution routes connect a declared SRGM application to scoped L1/L2 reports and distinct record/evidence status. Bounded witness discovery is separate from the certificate checker, which derives count and orientation without an expected count. The routes discharge different obligations; certificate verification does not replace representation, provenance, or population-evidence assessment. L3 requires separate evidence. Exact grammar, certificate schema and backend assumptions are specified in Supplement S7.}
\label{fig:audit-workflow}
\end{figure}

Algorithm 1 specifies the logical audit obligations. Individual execution routes discharge only their supported obligations and explicitly report components that are not assessed. Figure~\ref{fig:audit-workflow} summarizes the two routes, their evidential authorities and the separate population-evidence assessment.

\textbf{Algorithm 1. Audit a declared SRGM application.}

\begin{enumerate}
\def\labelenumi{\arabic{enumi}.}
\tightlist
\item
  Parse the contract, units/origin and supported scope; reject malformed
  input or retain conditional scope where required.
\item
  Check whole-horizon rate admissibility and attach its witness; keep
  observational adequacy separate.
\item
  Classify the selected mean via \(q_p\), strict horizon filtering and
  sign/orientation checks, retaining endpoint/noncrossing diagnostics
  and abstentions.
\item
  Apply the representation, numerical, semantic and provenance checks supported by the selected route; explicitly identify obligations not assessed by that route.
\item
  Assess whether the population target has the
  calibration/information/selection evidence required for
  interpretation.
\item
  Export L1/L2/L3 and record status, preserving sensitivities and all
  unresolved components.
\end{enumerate}

Validity is checked before shape; extending the horizon requires a new fitted classification and a separate inferential assessment.

\hypertarget{producer-checker-and-evaluation-references}{%
\subsection{Producer, checker and evaluation
references}\label{producer-checker-and-evaluation-references}}

Two execution routes have distinct authorities. The record-validation route constructs numerical records and validates them against out-of-band trusted contracts and frozen reference labels. Its structural gate, provenance checks and abstention semantics remain unchanged for replay; its results are not retrospectively promoted to interval certificates.

The certificate route first parses exact tagged values and checks admissibility. A bounded producer then searches for witnesses, and a separate checker verifies the resulting certificate before exporting a versioned record. The checker receives only the contract and certificate, never an expected count or manifest label. It reconstructs the canonical polynomial and coordinate map, recomputes the required enclosures, and checks signs, derivatives, complete coverage, multiplicity and open-horizon filtering. It derives count and orientation from the verified witnesses. The same semantics apply to registered and previously unseen supported inputs. Floating-point hints assist discovery only. Missing evidence or an exhausted budget does not establish invalidity. Literal context binding is checked separately from numerical equivalence.

The fixed verifier backend uses exact rational arithmetic and rigorous rational enclosures for logarithmic and exponential terms. This backend and the shared parser remain trusted implementation components. The checker neither imports the producer's classification nor accepts arbitrary enclosure callbacks. Derivatives of the canonical numerator certify local isolation; the physical-time derivative $q'$ is computed separately for robustness. The supplied contract must still represent the source model correctly. This route does not assess representation, and L3 remains unresolved. A standard-library CLI can recheck a certificate without fitting, a producer or manifest labels (Supplement S7 and S11). Here, exactness refers to the denotations of the supplied inputs and to the arithmetic or enclosure predicates. It does not imply that printed parameters equal the source authors' unrounded estimates.

The symbolic references serve evaluation purposes. They independently differentiate the rates, isolate polynomial roots and check the link from contract to contrast. Separate implementations do not establish independent human review or formal verification of the programs. The record-validation cache adapter reports unknown keys as unresolved and does not provide ground truth for the certificate route.

\begin{figure}[!t]
\begin{boxwithhead}{Box 1. Worked certificate: the archived constructed two-root case.}{

\textbf{Contract.} Rational-logistic, $A=100$, $p=c=r=1$, $\beta=100$, $k=1/10$; mean reset at $t_0=0$, time in days and physical rate in day$^{-1}$. Validity uses $[0,\log1000]$ and inflections its open interior. All scalar leaves are exact tagged rationals. The endpoint $T=\log1000$ is an exact real-valued time expression in the supported grammar. Under $x=100e^{-t}$, the closed validity horizon maps to $[1/10,100]$, with interior inflections restricted to $(1/10,100)$. The condition $k>0$ supplies the closed-horizon nonnegative-rate witness, and
\[
 P(x)=-1+\tfrac7{10}x-\tfrac1{100}x^2,\qquad q(t)=\frac{P(x)}{(1+x)^2}.
\]
\textbf{Local isolation.} The stored boxes are $J_1=[5/8,15/4]$ and $J_2=[u,v]$, where
\[
 u=\frac{10959451579617971}{175921860444160},\qquad
 v=\frac{13156275811914419}{175921860444160}.
\]
Their endpoint signs are $(-,+)$ and $(+,-)$. The checked derivative enclosures are $P'(J_1)\subset[5/8,11/16]$ and $P'(J_2)\subset[7/10-v/50,7/10-u/50]\subset(-\infty,0)$, giving exactly one simple zero per box. These are $x$-derivatives of $P$.

\textbf{Complement coverage.} The remaining pieces $[1/10,5/8]$, $[15/4,u]$ and $[v,100]$ have checked $P$-range enclosures $[-9301/10000,-145/256]$, $[95/64,45/4]$ and $[-31,P(v)]$, respectively, with $P(v)<0$. Together with the root boxes they cover the closed coordinate horizon without gaps; $q$ has the same sign as $P$.

\textbf{Checker output.} Producer-free rechecking returns \texttt{ACCEPTED}, diagnostic \texttt{CERTIFICATE\_VERIFIED}, validity \texttt{valid} and count $2$. In increasing physical time the orientations are $-\!\to\!+$, then $+\!\to\!-$: the decreasing coordinate map reverses box order and crossing direction. No endpoint or noncrossing zeros are reported. The count is derived from checked witnesses and coverage, not an expected-count input.

\textbf{Assurance and locator.} The result uses the fixed verifier backend (\texttt{aup08.fixed-fraction/1.0.0}; nine log/exp terms) under Section 3.4's trust boundary. It is not kernel-checked; representation is not assessed and L3 is unresolved. Inputs are \texttt{contract.json} and \texttt{certificate.json} under \path{supplementary/pipeline/replay_inputs/rational_two_roots/}; the recheck output is \path{paper/worked_certificate/verification.json}. Supplement S7 and S11 give the full semantics and CLI. The IBM record in Table~\ref{tab:4} retains \texttt{POINT\_SHAPE\_NUMERIC}.
}
\end{boxwithhead}
\end{figure}

\hypertarget{a-complete-worked-record-for-ibm-entry}{%
\subsection{A complete worked record for IBM
Entry}\label{a-complete-worked-record-for-ibm-entry}}

Table 4 presents the primary logistic record in readable form.
Parameters are shown in the original day units, rounded for
presentation; the accompanying JSON preserves the selected tuple, full
diagnostics, evaluation outputs and provenance. The record keeps
numerical eligibility, mathematical decisions and population evidence as
separate fields.

\begingroup

\begin{table}[!t]
\caption{IBM Entry primary logistic audit record.}\label{tab:4}
\centering
The full
archived record is \texttt{ibm\_entry\_logistic\_70\_\_free\_\_base} in
\texttt{e4/audit\_records.json}; this table is its
readable projection.
\small
\begin{tabular}{@{}
  >{\raggedright\arraybackslash}p{(\textwidth - 2\tabcolsep) * \real{0.2200}}
  >{\raggedright\arraybackslash}p{(\textwidth - 2\tabcolsep) * \real{0.7800}}@{}}
\toprule\noalign{}
\begin{minipage}[b]{\linewidth}\raggedright
Field
\end{minipage} & \begin{minipage}[b]{\linewidth}\raggedright
Recorded value and interpretation
\end{minipage} \\
\midrule

Data and source & 21 daily intervals; 46 failures; \cite{ref9}, Table V,
printed p.239; source explicitly declares origin zero \\
Contract & \(b(t)=c/(1+\beta e^{-rt})\), \(p=1\), \(m(0)=0\);
\(m(t)=K[1-e^{-B(t)}]\), \(B(t)=\int_0^t b(s)ds\) \\
Selected parameters & \(K=99.30137229\), \(c=0.055496779\) day\(^{-1}\),
\(r=0.170575743\) day\(^{-1}\), \(\beta=3.894960064\); free \(c,r\) \\
Training and selection & 14 intervals, 30 failures; base bounds;
training-likelihood/numerical-gate selection; L-BFGS-B \\
Numerical decision & \texttt{ACCEPTED\_NUMERIC}; no bound-hit flag;
rechecked training log likelihood \(-22.706727617\); maximum projected
score per count \(5.30\times10^{-8}\); local-probe improvement 0 \\
L1: mathematical validity & Valid on \([0,21]\): \(c,r,\beta>0\) imply
positive rate; NHPP and independent disjoint increments remain
observation assumptions \\
L2: training horizon & On \((0,14)\): convex mean, count 0; no roots,
orientations, boundary zeros or noncrossing zeros \\
L2: extended horizon & On \((0,21)\): count 1, root \(14.553874076\)
days, orientation \(+\to-\); no boundary/noncrossing zeros; root lies
beyond training \\
Representation and evidence & Elementary mean; point-shape status
\texttt{POINT\_SHAPE\_NUMERIC}; shape evidence numerically supported by
rule evaluation and the archived reference recheck \\
L3: population evidence & \texttt{INFERENCE\_HOLD}; no accepted
coverage/selective calibration for this history/horizon; no latent
population-root truth; no band transferred to the forecast horizon \\
Held-out evaluation & 7 intervals, 16 observed failures; predicted total
19.10310032; test NLL/interval 1.86924133; increment RMSE 1.77992787;
evaluation only \\
Evidence locator &
\path{srgm_fitting_policy_v1/results/final_decisions.json} record index
25; selected fit and numeric/shape rechecks retained in the E4 record \\
Provenance limits & Published-table transcription; automated
increment/total checks; original cited dataset and independent double
entry not checked; data license unspecified \\
\bottomrule
\end{tabular}
\end{table}
\endgroup

The permissible report is: \emph{The training-selected logistic mean is
convex on the 14-day horizon and has one fitted inflection on the 21-day
horizon. The latter is an extrapolated fitted feature; population-shape
evidence remains unresolved.} The numerical and L1/L2 fields justify
this fitted description, while the L3 field blocks its promotion to a
population conclusion. Complete record bytes, the data hash and all
source locators are preserved with Supplement S6.3.

\hypertarget{reproducible-execution}{%
\subsection{Reproducible execution}\label{reproducible-execution}}

The reproducibility archive provides a top-level quick profile for integrity checks, source-review reconciliation, certificate rechecking, pipeline tests, and a constructed two-root example. Separate entry points reproduce the E1--E4 analyses under their pinned dependencies. The quick profile does not rerun full fitting, population calibration, or Lean compilation. Supplement S6 and S11 distinguish fresh computation, archived-result aggregation, and proof checking; S12--S13 provide source traceability and electronic record locators.

\hypertarget{evaluation-design}{%
\section{Evaluation design}\label{evaluation-design}}

Table 5 links each claim to its supporting evidence and principal limit. The study designs below keep their denominators and evidence targets separate.

\hypertarget{e1-decision-correctness}{%
\subsection{E1 Decision
correctness}\label{e1-decision-correctness}}

E1 checks validity, inflection count, crossing orientation, endpoint and noncrossing lists, and eligible coordinates. Classical analytical labels and separate Decimal70 calculations assess constructed boundary cases. QUADPACK integrates the rate without using its primitive; DOP 853 solves the ODE without using the primitive or oracle. Fifteen configurations are run at three tolerances under prespecified numerical criteria. Generated rational cases are compared with Sturm counts, and interval inequalities undergo separate arithmetic checks. We report agreement together with conservative abstentions, unsupported inputs and unresolved reference predicates, retaining these outcomes in the denominators.

\hypertarget{e2-baselines-labels-and-benign-controls}{%
\subsection{E2 Baselines, labels and benign
controls}\label{e2-baselines-labels-and-benign-controls}}

The four computational baselines check different properties. N1 checks registered top-level schemas and types. N2 samples the rate sign. N3 counts Float64 curvature sign changes on a 2001-point grid. N4 checks internal consistency among count, shape and representation. Their union rejects a record whenever any baseline flags an error. These limited checks are neither complete audit workflows nor interval certificates. Their definitions remain frozen from the original campaign and are replayed unchanged against the final audit.

Two targets are separated. \textbf{Semantic/reference comparison}
assesses validity, mean shape and zero payloads under the record's own
declared contract, using independently computed truths from the frozen
cache. \textbf{Provenance integrity} asks whether the selected field
matches its trusted original contract. Changing a parameter may create a
mathematically valid new application while violating original
provenance. Reference labels and rejection reasons are separate axes, as
formalized below.

\textbf{Record-rejection measure.} Let \(M\) be the 840 reference-labelled mathematical-disagreement records and let \(R_i=1\) denote rejection of record \(i\). We report
\[\mathrm{RR}_M=\frac{\sum_{i\in M}R_i}{|M|}.\]
This measure includes rejection for mathematical, structural, provenance or evidence obligations. It does not establish that the executed failure identifies the reference-defined mathematical defect: a wrong-count record rejected only for a bad code hash still contributes to rejection. Obligation-matched diagnosis is not evaluated; baselines and ablation use record rejection throughout.

The cache-comparison adapter applies Decimal90 coordinate tolerances capped by endpoint distances. Before the new-seed analysis, it exactly reproduces all 10,440 stored comparisons in the original independent-oracle corpus. It derives no new symbolic roots. Missing cache keys and unresolved original predicates remain reference unresolved. Confirmation mutations use the same frozen fixture contracts, so they do not validate held-out families. Cached labels never use a record's expected-label field.

Benign controls preserve semantics by adding a presentation annotation
or reordering reasoning metadata in 261 records. Separately, 1,220
random attempts leave the selected contract field unchanged. These
narrow transformations do not establish acceptance of every
mathematically equivalent reparameterization. All false rejections of
these controls are reported. A mathematically consistent record is not
automatically benign with respect to provenance, evidence, schema or
route output.

\textbf{E2-B benign semantic controls.} We execute supported
representation controls separately from equivalent-application controls
through the frozen \texttt{verify\_json} interface. B1 changes raw JSON
key order/indentation/whitespace, with parsed-value equality. B2 uses
unreduced rational strings, explicit-e0 decimals and normalized
scientific notation, with an independent exact integer/regex parser.
Schema types remain unchanged; record and trusted lexical context are
rebound together. B3 applies \(t'=a t\), \(r'=r/a\) and \(c'=c/a\) where
applicable, with \(a=24\) and \(1/24\), transporting horizons,
root/sample coordinates and witnesses. The identity
\(b_{\rm new}(at)=b(t)/a\) implies mean equality and
\(q_{{\rm new},p}(at)=q_p(t)/a^2\), preserving strict crossings and
their orientation. Independent Decimal100 checks and mean-sample
comparisons establish numerical relation witnesses before audit. Generic
source/code hashes stay unchanged; new execution logs are recorded. The
API exposes numeric contracts, not an automatic
unit/tolerance-conversion service: fixed validator tolerance floors
remain unchanged, and coordinate transport uses a separately scaled
error budget. Eligibility and status-preserving strata are explicit.
Companion negatives retain the old trusted context; their provenance
rejection is expected. Full manifests, capability checks and traces are
in Supplement S3.6 and the E2-B artifact.

\textbf{Post-freeze composite-mutation challenge.} We locked a separate protocol before applying one coupled shape-profile mutation to each eligible base route record. Eligibility requires a resolved, valid and reference-consistent base record. A base with zero crossings receives a coherent one-crossing payload at the horizon midpoint. A base with one or two crossings receives a coherent zero-crossing payload. Each mutation changes the shape, count and root fields together, while preserving the contract, context and numerical records. This operator family was absent from the original single-field generator and was selected after the original results. It therefore tests neither held-out cases nor held-out rate families. The validator and N1--N4 remain frozen, and the existing independent cache labels each changed payload. The evaluation addendum retains the protocol, requests, exclusions, full records and traces.

\textbf{Decision-rule ablation.} To separate contributions within the
audit, we retain the front-end gate and partition all semantic failure
codes into five disjoint groups: D (domain/admissibility and scope
gates), S (shape/zero/horizon rules), E (evidence semantics), P (trusted
provenance/locators) and R (representation/numerical-route checks). We
compare front-end-only, front-end plus each group alone, full audit
minus each group and full audit on the same frozen confirmation
attempts. For front-end-admissible records, rejection is the disjunction
of retained failure codes; malformed inputs retain the schema rejection.
Every observed code is mapped once, and the full configuration
reproduces every saved audit outcome. This is an output decision-rule
ablation: computations, branches and the front-end gate remain
unchanged. Single-group policies are not deployable complete audits.
Group definitions were fixed before this post hoc analysis; paired loss
intervals use 2,000 resamples of 87 case clusters, seed 20261019.
Exclusive rejection losses measure necessity within the saved trace, not
independent causal contributions or performance effects. All per-attempt
outcomes are retained in the ablation addendum.

\hypertarget{e3-inference-protocol-and-generator-truth}{%
\subsection{E3 Inference protocol and generator
truth}\label{e3-inference-protocol-and-generator-truth}}

Given cumulative counts \(C_i\), fitting uses disjoint increments
\(y_i=C_i-C_{i-1}\), assumed independent Poisson under the NHPP model.
Cumulative observations are not independent samples. At fixed \(p=1\),
\(m=Kh_\theta\), and \(d_i=h_\theta(t_i)-h_\theta(t_{i-1})\),

\[\ell(K,\theta)=\sum_i[y_i\log(Kd_i)-Kd_i-\log(y_i!)],\qquad
\widehat K(\theta)=\frac{\sum_i y_i}{\sum_i d_i}.\]

The locked bounded fitting policy uses training data only and distinguishes numerical eligibility from optimizer termination. Failed draws are retained without redrawing. Expanded bounds and the source constraint \(c=r\) are assessed as sensitivities. Test scores never select configurations. Supplement S3/S5 gives the optimization budgets, thresholds and diagnostics.

\textbf{Why this inference procedure is audited.} E3 uses a common
sign-preserving curvature contrast to examine the step from a fitted
curve to a population-shape claim across the three rate families. Let
\(\mathcal H_\theta\) denote the family-specific contrast in normalized
time, sign-equivalent to \(q_p\) (Supplement S4). The bounded transform
\(\psi=\mathcal H_\theta/(1+|\mathcal H_\theta|)\) preserves zeros and
signs and permits a shared sup-norm band with deterministic zero/count
projection. This makes two obligations separately testable: whether
projection is sound conditional on containment, and whether the
fitted-data procedure supplies the claimed containment probability.
Joint parametric refits accommodate the same observation model and
constrained fitting policy used elsewhere in the audit. The procedure is
a working audit target, not an inferential innovation or a
recommendation established by this study. Its design rationale is the
transparent separation of deterministic and probabilistic obligations;
calibration assesses the latter rather than assuming it.

The evaluated procedure uses 100 joint refits and a 1001-point grid, supplemented by analytical local Lipschitz bounds. Each nonaccepted draw contributes infinity, and the 91st order statistic defines the radius \(E\). Conditional on \(\|\psi_0-\widehat\psi\|_\infty\le E\), exact inversion with the stated polynomial and sign restrictions contains the true zeros and count. This deterministic result does not establish the probability of the containment event. We therefore report band containment, all-zero-region containment, count-set containment, width/T, singleton yield and the false frequency among singleton decisions separately. E3 applies to this transform, radius rule, refit budget, bounded fitting policy and tested boundary/count design. Supplement S4 explores parameter-set projection, profile-likelihood and Bayesian alternatives; E3 does not establish their calibration.

Independent confirmation has 20 cells: ten mechanisms at expected counts
100/500, 50 histories per cell and 50 disjoint increments per history.
Mechanisms distinguish endpoints, folds and rate-admissibility
boundaries; a rescaled delayed control is not a new difficulty class.
Cell-specific precision is retained without simultaneous adjustment.

\textbf{Ideal-fold and rounded-generator targets.} The ideal fixture has
\(D=0\) and no sign-changing zero; the binary-rounded generator has a
tiny positive discriminant and two interior crossings. Both targets are
retained. The count-only sensitivity in Supplement S4 re-evaluates the
frozen histories without refitting; high-precision mean differences are
not interval-certified bounds.

\hypertarget{e4-and-common-reporting-policy}{%
\subsection{E4 and common reporting
policy}\label{e4-and-common-reporting-policy}}

TROPICO/wireless tables {[}7, p.503{]} and IBM Entry {[}9, p.239{]}
supply 81/461 weekly, 51/203 weekly and 21/46 daily interval/failure
totals. Primary training uses floor(70\%) intervals with 60\%/80\%
sensitivities. Origin zero is explicit for IBM and a convention for the
others. Automated increment/total checks retain source provenance, while
independent double entry and original cited-data/rights checks remain
incomplete. Bands stay on the training horizon; unknown population roots
prevent empirical latent-root coverage evaluation.

All studies retain requested and accepted denominators, mathematical and
malformed/integrity labels, and unresolved outcomes \cite{ref29}. The
baseline and ablation analyses use the same frozen seed-20261016
mutation corpus, with no validator tuning during analysis. Cluster
intervals use 2,000 resamples of 87 input cases, preserving
route/mutation siblings; insufficient-cluster and degenerate intervals
are omitted. Exceptions and internal abstentions are not detections.
Supplement S3/S6 specifies final evidence, replay routes, retained
misses and unresolved draws.

\subsection{Locked certificate campaigns and source-point audit}
The primary certificate campaign contains 180 synthetic requests. A separately locked extension adds 90 requests to broaden parameter coverage. After adjudication, the printed-point audit evaluates 30 supported contracts: 16 proposed new-source points and separate replay, previous-source and mean-only groups. These purposive designs are neither population samples nor a systematic review. We assess correctness, yield, abstention and cost using a symbolic differentiation/Sturm reference, default repeats, tight-budget runs, hints-off runs, a 4097-node numerical grid, margin profiles and producer-free verification. All requested denominators and unresolved cases are retained. The operational pre-run locks are not external preregistration. Supplement S9--S10 reports protocols, dependencies, failures and source eligibility.

\noindent\textbf{Source-to-contract evidence.}
Two human reviewers each assessed 13 fields for all 30 printed-point contracts, yielding 390 paired assessments. They agreed on 326 matches and ten unresolved items; labels differed on 54 assessments. Documented reasons concerned review scope (30), audit horizon versus unknown estimation span (6), fit-metric replay and rounding (5), clock and replay availability (8), unit/horizon/lineage reservations (4), and mean mapping versus a disputed printed rate (1). These differing labels are not a count of transcription errors. Original labels and remaining reservations are retained in Supplement S12--S13.

The evidence chain links source locators and printed literals to the unchanged contract, literal input digest, certificate digest, archived reference and fresh verifier result. We recompute all 30 input digests. These checks establish content binding and traceability. They do not establish publication authenticity or validate the shared parser and arithmetic backend.

\hypertarget{results}{%
\section{Results}\label{results}}

We answer RQ2 in two parts. Section 6.2 reports record rejection and control outcomes for the record-validation route. Section 6.5 reports correctness, yield and rechecking for the certificate route. Table 5 distinguishes the evidence and interpretation limits of the two routes.

\begin{table}[!t]
\caption{Claims, supporting evidence and interpretation limits.}\label{tab:5}
\centering
\small
\begin{tabular}{@{}>{\raggedright\arraybackslash}p{0.23\linewidth}>{\raggedright\arraybackslash}p{0.43\linewidth}>{\raggedright\arraybackslash}p{\dimexpr0.34\linewidth-4\tabcolsep\relax}@{}}
\toprule
Claim & Supporting evidence & Interpretation limit\\\midrule
Validity and fitted shape (RQ1) & Classification, certificate soundness, separated-simple-root stability and reference-search completeness (Section 3; S2, S7--S8). & Conditional mathematical guarantees; concrete parser/backend/checker are outside the Lean proof.\\\addlinespace[3pt]
Certificate execution (RQ2) & Primary 169/174, extension 87/87 and printed-point 30/30 admissible inputs certified; rechecking and budget runs (Section 6.5; S9--S10). & Yield depends on supported encodings and budgets; no population error-rate guarantee.\\\addlinespace[3pt]
Record rejection (RQ2) & Frozen corpus 840/840 versus union 646/840; composite challenge 234/234 versus 186/234; 424 integrity misses retained (Section 6.2; S3; evaluation addendum). & Rejection is not obligation-matched diagnosis; overall false-rejection rates beyond the specified control classes and semantic-impact rates for retained changes are not estimated.\\\addlinespace[3pt]
Limits of the working inference procedure (RQ3) & 11 false outputs among 22 singletons from 200 delayed-boundary histories; uninformative sets and band failures (Section 6.3; S4). & Results concern this band/projection procedure, fitting policy and tested design.\\\addlinespace[3pt]
Source-linked reporting & Three histories and 162 fitted records; two reviews of 30 contracts across 13 fields (Sections 5.5, 6.4--6.5; S5, S10, S12--S13). & Source reservations remain; shared histories are not independent datasets and population roots are unknown.\\
\bottomrule
\end{tabular}
\end{table}

\hypertarget{rq1-obligations-guarantees-and-abstention}{%
\subsection{RQ1 Obligations, guarantees and
abstention}\label{rq1-obligations-guarantees-and-abstention}}

\textbf{Answer to RQ1.} Validity requires whole-horizon rate
admissibility under the declared dynamics and origin; fitted shape
requires complete interior-zero classification with crossing and
endpoint checks. The certificate theorem guarantees soundness only when
its root-box, derivative and complement-coverage premises hold. For the certificate route, missing or unresolved required witnesses cause abstention on the corresponding certified decision. The record-validation route may still report a numerically supported fitted classification under its own checks. Sections 2--3 establish these obligations and
conditional guarantees; E1 evaluates their execution on supported cases. Certificate-route correctness, yield, budget sensitivity and rechecking are evaluated separately in Section 6.5 under the design in Section 5.5.

All 80 classification checks and 45 representation runs meet their
criteria. Maximum absolute ODE mean error decreases from
\(6.88\times 10^{-6}\) at coarse tolerance to \(9.75\times 10^{-10}\) at
the finest; scaled Float64 mean error versus Decimal70 is
\(2.70\times 10^{-14}\). Eligible root error normalized by \(1+T-t_0\)
is \(2.43\times 10^{-13}\). Endpoint/double roots are excluded from that
coordinate metric. Thirty-two perturbation regressions and a retained
oscillatory counterexample show why function sup-norm control alone does not
ensure uniqueness.

Of 261 base records, 234 valid-domain decisions match independent
reference shape/payload labels, nine invalid-rate decisions and 15
conditional/out-of-scope decisions match their appropriate labels, and
three valid near-endpoint cases conservatively abstain. Thus 258
comparisons agree and three abstain; all 261 records pass the final
record audit, including 19 correctly scoped unresolved evidence outputs.
Record acceptance and mathematical resolution have different
denominators.

The interval reference certifies 82 of 160 cached input specifications,
marks 24 outside scope and retains 54 unresolved. Two hundred generated
rational cases agree with Sturm counts; 2,852 nonzero sign inequalities
pass separate Arb/FLINT checks. These results provide finite-case agreement and explicit abstention evidence for the obligations in Sections 2--3, within the trust boundaries of Section 3.4.

\hypertarget{rq2-record-checking-and-added-rejection-capability}{%
\subsection{RQ2 Record checking and added rejection
capability}\label{rq2-record-checking-and-added-rejection-capability}}

\textbf{Answer to the record-validation component of RQ2.} On the frozen confirmation corpus, the audit
rejects all reference-labelled mathematical disagreements and malformed
decision payloads, rejects more labelled records than the union of the
four computational baselines, and accepts the defined benign and
unchanged controls. This result concerns labelled decision errors and
specified controls; provenance-integrity sensitivity is a separate
measure with residual misses.

The confirmation annotation divides the 10,440 attempts into five groups: 840 mathematical decision or zero-payload disagreements; 1,761 malformed decision payloads; 7,510 reference-consistent decisions; 97 conservative shape abstentions; and 232 reference-unresolved inputs. The last two groups remain in the report. The first three provide definite comparisons for 10,111/10,440 attempts. Annotation uses each record's declared contract and the frozen independent cache. Its denominator differs from the 9,220 records whose selected field changed.

\begin{table}[!t]
\caption{Final audit versus frozen computational baselines on
seed 20261016.}\label{tab:6}
\centering
Columns report record rejection within labelled subsets
or false rejection of defined controls (Section 5.2).
\small
\begin{tabular}{@{}
  >{\raggedright\arraybackslash}p{(\textwidth - 10\tabcolsep) * \real{0.1304}}
  >{\raggedleft\arraybackslash}p{(\textwidth - 10\tabcolsep) * \real{0.1739}}
  >{\raggedleft\arraybackslash}p{(\textwidth - 10\tabcolsep) * \real{0.1739}}
  >{\raggedleft\arraybackslash}p{(\textwidth - 10\tabcolsep) * \real{0.1739}}
  >{\raggedleft\arraybackslash}p{(\textwidth - 10\tabcolsep) * \real{0.1739}}
  >{\raggedleft\arraybackslash}p{(\textwidth - 10\tabcolsep) * \real{0.1739}}@{}}
\toprule\noalign{}
\begin{minipage}[b]{\linewidth}\raggedright
Checker
\end{minipage} & \begin{minipage}[b]{\linewidth}\raggedleft
Math-disagreement records rejected /840
\end{minipage} & \begin{minipage}[b]{\linewidth}\raggedleft
Malformed /1,761
\end{minipage} & \begin{minipage}[b]{\linewidth}\raggedleft
Unchanged rejected /1,220
\end{minipage} & \begin{minipage}[b]{\linewidth}\raggedleft
Metadata controls rejected /261
\end{minipage} & \begin{minipage}[b]{\linewidth}\raggedleft
Integrity detected /9,220
\end{minipage} \\
\midrule\noalign{}

N1 Types & 48 & 1,231 & 0 & 0 & 4,164 \\
N2 Rate sampling & 115 & 278 & 54 & 9 & 860 \\
N3 Curvature grid & 466 & 509 & 341 & 72 & 2,951 \\
N4 Consistency & 331 & 472 & 0 & 0 & 1,490 \\
Union N1--N4 & 646 & 1,448 & 341 & 72 & 6,238 \\
Final audit & 840 & 1,761 & 0 & 0 & 8,796 \\
\bottomrule
\end{tabular}
\end{table}

The union rejects 76.90\% of reference-labelled
mathematical-disagreement records (95\% case-cluster CI: 72.62--81.34\%), versus \(\mathrm{RR}_M=840/840=100\%\) for the
audit. Its malformed-payload rate is 82.23\% versus all 1,761 rejected.
Degenerate perfect-rejection intervals are omitted rather than treated
as population guarantees. The audit adds explicit domain/horizon/zero
semantics and record obligations that the limited baselines do not
jointly establish. Their false rejection of some valid
conditional/abstention controls also shows why grid flags cannot replace
the record semantics. Metric interpretation follows Section 5.2.

The audit accepts all 261 benign metadata transformations and all 1,220
unchanged attempts. Acceptance beyond the transformations specified in
E2-B remains unevaluated. Among 7,510 mathematically
reference-consistent records it rejects 5,876; these are not
automatically mathematical false positives because reference consistency
does not validate complete schema, provenance, representation or
evidence. Among 97 conservative shape abstentions, 87 also violate other
record obligations; ten complete records remain accepted unresolved.
Rejection of a malformed/reference-unresolved input does not resolve its
mathematical model.

Integrity is reported separately. The audit rejects 8,796 of 9,220 changed-field records (95.40\%; case-cluster 95\% CI 94.87--95.89\%), compared with 6,238/9,220 (67.66\%) for the baseline union. The 424 retained misses concern parameter-domain fields (123), predicate witnesses (94), initial conditions (92), representation fields (45), horizons (29), theorem/derivation/source locators (30), implementation status (8), component evidence (2), and a numerical result (1). This is a field-level integrity breakdown, not an obligation-matched mathematical-diagnosis result. These retained changed-field records are misses under the strict integrity metric; that classification does not establish that each change is a harmful semantic error. Full records and field counts are in Supplement S3.3.

Compared with the baseline union, the audit rejects an additional 194/840 mathematical-disagreement records and 313/1,761 malformed-payload records. It also avoids the union's 72 rejections of defined benign controls and 341 rejections of unchanged attempts. These gains apply to the labelled corpus. The audit still retains 424 changed-field records, so it is not exhaustive under the strict integrity metric. The semantic effects of those changes require separate assessment. Estimating an overall mathematical false-positive rate would also require obligation-specific adjudication of the reference-consistent and reference-unresolved groups.

\hypertarget{isolating-added-value-by-decision-rule-ablation}{%
\subsubsection{Isolating added value by decision-rule
ablation}\label{isolating-added-value-by-decision-rule-ablation}}

Table 7 reports record rejection under retained rule groups. The
front-end gate alone rejects 127/840 mathematical-disagreement records
and all 1,761 malformed decision payloads. Schema plus S rejects
674/840. Removing S loses 541/840 rejections (64.40 percentage points;
cluster CI 59.82--68.63): shape-rule failures are necessary for
rejecting these records within the saved trace. Removing D loses 62
rejections. Provenance rules uniquely reject 104 records within the
mathematical-disagreement category: full minus P rejects 736/840. These
saved-trace losses measure rejection necessity (Section 5.2); shared
failures make them nonadditive.

\begin{table}[!t]
\caption{Fixed-front-end decision-rule ablation on the same
confirmation corpus.}\label{tab:7}
\centering
Loss columns are paired losses relative to the
full audit. All configurations reject 1,761/1,761 malformed decision
payloads and reject zero of the 1,220 unchanged attempts. D:
domain/scope; S: shape/zeros/horizon; E: evidence; P: provenance; R:
representation. Full outcome details and intervals are archived.
\small
\begin{tabular}{@{}
  >{\raggedright\arraybackslash}p{(\textwidth - 8\tabcolsep) * \real{0.1579}}
  >{\raggedleft\arraybackslash}p{(\textwidth - 8\tabcolsep) * \real{0.2105}}
  >{\raggedleft\arraybackslash}p{(\textwidth - 8\tabcolsep) * \real{0.2105}}
  >{\raggedleft\arraybackslash}p{(\textwidth - 8\tabcolsep) * \real{0.2105}}
  >{\raggedleft\arraybackslash}p{(\textwidth - 8\tabcolsep) * \real{0.2105}}@{}}
\toprule\noalign{}
\begin{minipage}[b]{\linewidth}\raggedright
Configuration
\end{minipage} & \begin{minipage}[b]{\linewidth}\raggedleft
Math-disagreement records rejected /840
\end{minipage} & \begin{minipage}[b]{\linewidth}\raggedleft
Rejection loss in math-disagreement records
\end{minipage} & \begin{minipage}[b]{\linewidth}\raggedleft
Integrity detected /9,220
\end{minipage} & \begin{minipage}[b]{\linewidth}\raggedleft
Integrity loss
\end{minipage} \\
\midrule\noalign{}

Front-end only & 127 & 713 & 5,908 & 2,888 \\
Front-end + D & 189 & 651 & 6,200 & 2,596 \\
Front-end + S & 674 & 166 & 6,455 & 2,341 \\
Front-end + E & 133 & 707 & 6,173 & 2,623 \\
Front-end + P & 231 & 609 & 7,165 & 1,631 \\
Front-end + R & 127 & 713 & 6,441 & 2,355 \\
Full minus D & 778 & 62 & 8,504 & 292 \\
Full minus S & 299 & 541 & 8,255 & 541 \\
Full minus E & 840 & 0 & 8,537 & 259 \\
Full minus P & 736 & 104 & 7,539 & 1,257 \\
Full minus R & 840 & 0 & 8,263 & 533 \\
Full audit & 840 & 0 & 8,796 & 0 \\
\bottomrule
\end{tabular}
\end{table}

E and R have zero exclusive rejection gain in the
mathematical-disagreement category on this corpus, but their removal
loses 259 and 533 changed-field detections, respectively. Removing P
loses 1,257 integrity detections. Thus evidence semantics and
representation checks add integrity capability without an observed
exclusive gain on the mathematical-disagreement record-rejection metric.
All policies inherit acceptance of the 261 frozen benign metadata
controls because they retain subsets of the full audit's rejection
rules; this is a logical consequence, not a new execution test or a
broader benign-equivalence study. The ablation isolates group necessity
conditional on the retained front-end and saved trace. It does not
evaluate schema removal, alternative execution paths or a new error
distribution.

\hypertarget{supported-benign-semantic-transformations}{%
\subsubsection{Supported benign semantic
transformations}\label{supported-benign-semantic-transformations}}

Table 8 reports actual executions rather than saved-trace extrapolation.
B1 and B2 each preserve all 783 expected outcomes across 87 underlying
cases; B3 preserves 498 across 83 cases. There are no
equivalence-witness failures, status mismatches, unexpected rejections,
accepted relation violations or internal errors. The 24 ineligible B3
controls are retained with scope reasons. These are route/transformation
siblings, not independent new applications, and family rates are not
pooled.

\begin{table}[!t]
\caption{E2-B controls through the frozen executable interface.}\label{tab:8}
\centering
All eligible controls were executed. Status preservation includes
accepted-unresolved records; it does not imply validity of every
represented application. Rejections in the last two rows are expected
provenance negatives, not benign-control failures.
\small
\begin{tabular}{@{}
  >{\raggedright\arraybackslash}p{(\textwidth - 10\tabcolsep) * \real{0.1304}}
  >{\raggedleft\arraybackslash}p{(\textwidth - 10\tabcolsep) * \real{0.1739}}
  >{\raggedleft\arraybackslash}p{(\textwidth - 10\tabcolsep) * \real{0.1739}}
  >{\raggedleft\arraybackslash}p{(\textwidth - 10\tabcolsep) * \real{0.1739}}
  >{\raggedleft\arraybackslash}p{(\textwidth - 10\tabcolsep) * \real{0.1739}}
  >{\raggedleft\arraybackslash}p{(\textwidth - 10\tabcolsep) * \real{0.1739}}@{}}
\toprule\noalign{}
\begin{minipage}[b]{\linewidth}\raggedright
Control class
\end{minipage} & \begin{minipage}[b]{\linewidth}\raggedleft
Executed
\end{minipage} & \begin{minipage}[b]{\linewidth}\raggedleft
Expected status preserved
\end{minipage} & \begin{minipage}[b]{\linewidth}\raggedleft
Unexpected rejection
\end{minipage} & \begin{minipage}[b]{\linewidth}\raggedleft
Accepted relation violation
\end{minipage} & \begin{minipage}[b]{\linewidth}\raggedleft
Internal error
\end{minipage} \\
\midrule\noalign{}
B1 Serialization & 783 & 783 & 0 & 0 & 0 \\
B2 Exact value & 783 & 783 & 0 & 0 & 0 \\
B3 Clock covariance & 498 & 498 & 0 & 0 & 0 \\
B2 old-context companion & 783 & 783 expected rejections & n/a & n/a &
0 \\
B3 old-context companion & 498 & 498 expected rejections & n/a & n/a &
0 \\
\bottomrule
\end{tabular}
\end{table}

B3 includes 474 valid-application controls, 18 invalid-rate status
controls and six empty-horizon conditional-status controls. Valid
near-endpoint shape abstentions and solver-evidence-unresolved records
retain their distinct statuses. Four unsupported source cases contribute
24 nonexecuted controls. The maximum Decimal covariance residual is
\(9.0\times10^{-99}\); independently checked scaled mean-sample error is
at most \(1.822\times10^{-9}\), below the declared \(10^{-7}\) gate. B2
positive controls test exact-value representation preservation when both
record and trusted lexical context are rebound. Its old-context
companions test literal provenance binding: rejection is expected
despite numerical equivalence, and is neither a mathematical-error
diagnosis nor a false rejection under the declared contract. B3
companions test changed numerical contracts paired with the original
context. The observed invariance is limited to the specified
representations and two clock factors; no unit-aware tolerance API or
general equivalence guarantee follows.

\subsubsection{Post-freeze composite-mutation challenge}
Of the 261 base route records, 234 from 78 input cases meet the locked eligibility conditions. All 27 exclusions are retained. The audit accepts all 234 unmodified base controls, while the independent cache identifies a wrong count in every mutated payload. The audit rejects 234/234 mutations. The N1--N4 union rejects 186/234: 45/63 delayed, 30/42 logistic and 111/129 rational-logistic records. N1 and N4 reject none, and no payload is malformed. The additional rejections therefore cannot be attributed to top-level schema checks or internal count/list consistency alone. This post-freeze challenge tests an additional operator family on existing fixtures and references. It does not provide independent model-family validation or estimate diagnosis rates. A fresh replay also reproduces all 10,440 original audit outcomes and failure-code sets with the unchanged validator.

\hypertarget{rq3-calibration-and-limits-of-population-shape-inference}{%
\subsection{RQ3 Calibration limits of the evaluated inference
procedure}\label{rq3-calibration-and-limits-of-population-shape-inference}}

\textbf{Answer to RQ3.} For the evaluated band/projection procedure, correct classification of the selected curve
leaves three obstacles to population-shape claims: fitted counts can
differ from generating counts, projected sets can remain uninformative,
and nominal or selective error control can fail. E3 measures these
obstacles separately; numerical fitting eligibility is not an
inferential guarantee.

All 1,000 original confirmation fits are numerically accepted; 99,996
nested refits are accepted and four remain unresolved under the fixed
budget. Fifteen accepted termination warnings are retained.
Seed/gate/shape/inference checks cover 1,000 groups, with a fixed
40-decision fresh-worker subset in the same runtime, not complete
independent-machine replication.

Under the rounded-generator target, count disagreements decrease from
355 to 336; only the two rational fold cells change target count from
zero to two. Count-set containment remains 989/1,000 and false
singletons remain 11 (Supplement S4).

Three results constrain population interpretation. First, a post hoc
comparison finds 355/1,000 fitted inflection counts different from
ideal-convention targets, despite correct classification of the selected
curves. Second, every logistic endpoint history retains \(\{0,1\}\) and
every rational-fold history retains \(\{0,1,2\}\), so frequent count
containment supplies little discrimination. Third, 11 of 22 singleton
outputs in 200 delayed horizon-boundary histories are false; reporting
only 11 errors among 1,000 requested histories would conceal that
selective error. The pooled 50\% is descriptive across heterogeneous
cells, not a common-binomial guarantee.

\begin{figure}[!t]
\centering
\includegraphics[width=0.95\textwidth,keepaspectratio]{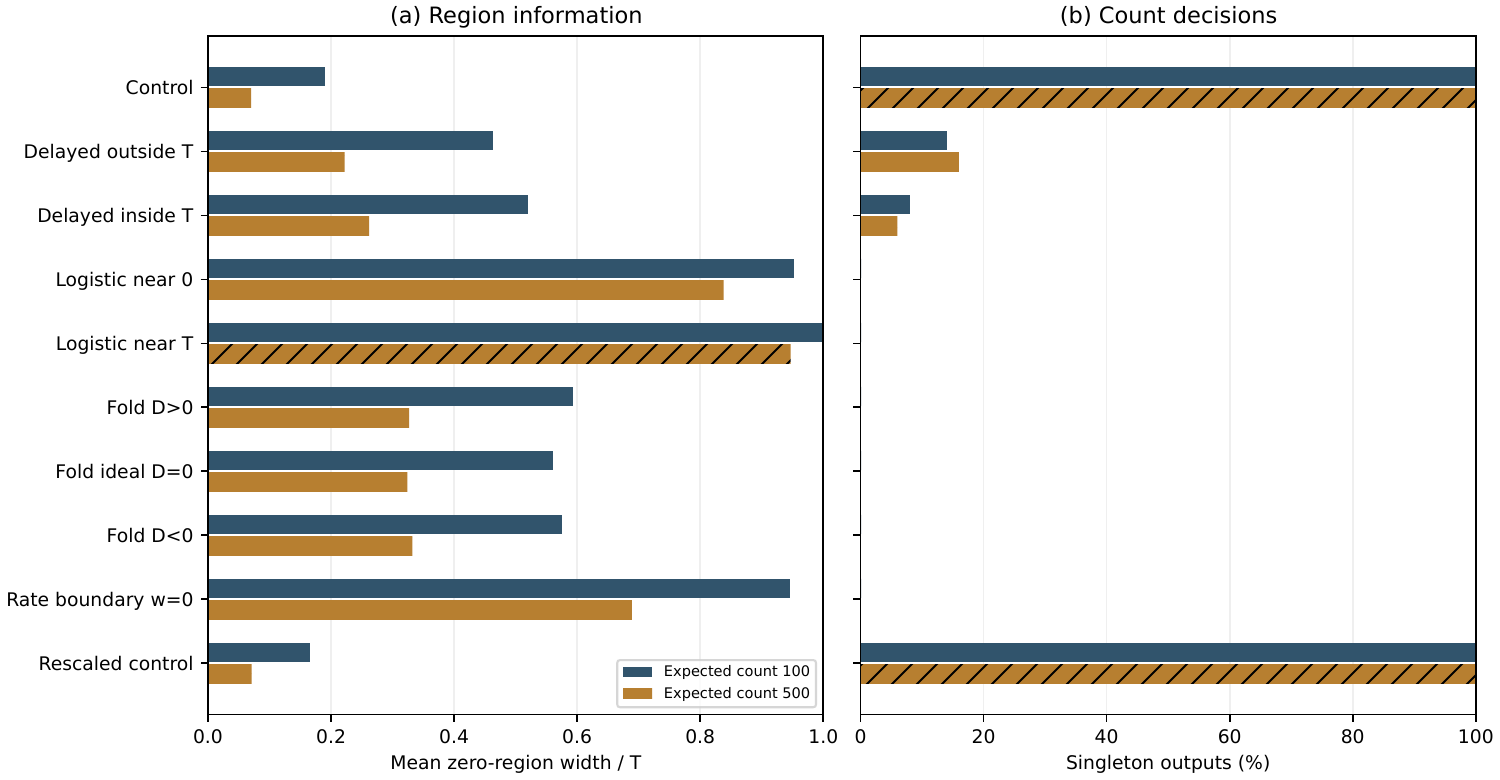}
\caption{Information and singleton yield in the frozen confirmation cohort. Panel (a) reports mean normalized zero-region width; panel (b) reports singleton-output percentages. Each bar summarizes 50 histories in one mechanism--expected-count cell. The two expected counts are 100 and 500. Higher expected count can contract zero regions while count ambiguity remains. Singleton yield alone does not establish correctness among selected outputs; selective correctness is reported in Table 9.}
\end{figure}

\begin{table}[!t]
\caption{Inference limits and reporting consequences.}\label{tab:9}
\centering
\small
\begin{tabular}{@{}
  >{\raggedright\arraybackslash}p{(\textwidth - 4\tabcolsep) * \real{0.3333}}
  >{\raggedright\arraybackslash}p{(\textwidth - 4\tabcolsep) * \real{0.3333}}
  >{\raggedright\arraybackslash}p{(\textwidth - 4\tabcolsep) * \real{0.3333}}@{}}
\toprule\noalign{}
\begin{minipage}[b]{\linewidth}\raggedright
Setting
\end{minipage} & \begin{minipage}[b]{\linewidth}\raggedright
Observed evidence
\end{minipage} & \begin{minipage}[b]{\linewidth}\raggedright
Permitted interpretation
\end{minipage} \\
\midrule\noalign{}

Logistic endpoint cells & 200 histories; all reported count sets equal \(\{0,1\}\);
normalized width 0.838--0.999 & No precise population count despite containment \\
Rational fold cells & 300 histories; all reported count sets equal \(\{0,1,2\}\) & Outer
sets retain all supported counts \\
Delayed horizon boundary & 22 singletons in 200 histories; 11 false &
Report selective error and abstention yield \\
Rate-admissibility boundary & Band 29/50 (58\%) at expected count 100 &
No nominal uniform-band claim \\
Published histories & Determinate fitted curves, unknown population
roots & Report fitted horizon features; keep L3 unresolved \\
\bottomrule
\end{tabular}
\end{table}

Band containment also falls to 78--80\% in several cells; high projected
containment does not repair the missing band probability premise. Each
cell has 50 histories, two-point proportion resolution and MC standard
error about 4.2 percentage points near 90\%. The current procedure is
neither generally calibrated 90\% confidence nor a validated
selective-shape rule. These failures limit the specified band/projection
procedure under the tested fitting policy and calibration design.

\hypertarget{rq3-reporting-consequences-on-published-histories}{%
\subsection{RQ3 Reporting consequences on published
histories}\label{rq3-reporting-consequences-on-published-histories}}

All 162 records passed the declared numerical and admissibility checks. Population-shape evidence remained unresolved for all three histories; the following results therefore describe selected fitted means. The records comprise 108 free and 54 source-constrained configurations, with 12 primary fits including the Goel--Okumoto (GO) constant-rate reference.

For TROPICO, primary training/full counts are 0/0 for GO, 1/1 for
delayed/logistic and 2/2 for rational. The rational fit retains a bound flag. Wireless gives 0/0 for GO and 1/1 for
the other families; logistic/rational retain bound flags. IBM gives 0/0 for GO, 1/1 delayed, 0/1 logistic and 2/2
rational. The logistic horizon change is the worked example; IBM GO
retains amplitude/limiting-profile diagnostics without implying a finite
unconstrained MLE. Full roots, orientations, forecasts and provenance
are in Supplement S5.

Bound expansion changes full-horizon counts in 3/36 comparisons and
forecasts beyond a declared 5\% diagnostic threshold in 6/36. Source
restriction \(c=r\) changes counts in 6/18 and forecasts beyond 5\% in
17/18. Count stability is not forecast stability, and these diagnostic
thresholds are not significance tests. Test scores evaluate
training-selected fits, never select the configuration or establish
population shape. E4 adds reportable distinctions and exclusions; latent-root coverage cannot be estimated from these histories.

\subsection{Certificate-route correctness, yield and cost}
\label{additional-certified-evidence}
\textbf{Answer to the certificate-route component of RQ2.} The supported campaigns show reference agreement of certified assertions, budget-dependent certificate yield and successful producer-free rechecking. These execution results complement the conditional soundness guarantees in Section 3; they do not replace the record-validation results in Section 6.2.

\begin{table}[!t]
\caption{Certificate-route outcomes with separate cohort denominators.}\label{tab:10}
\centering
\small
\begin{tabular}{@{}p{5.8cm}rrrr@{}}
\toprule Default-run cohort & Requested & Admissible & Certified & Abstain valid\\\midrule
Primary synthetic &180&174&169&5\\
Post-primary parameter extension &90&87&87&0\\
Proposed new-source points &16&16&16&0\\
Prior-point replay &4&4&4&0\\
Previously cited source extensions &8&8&8&0\\
Deterministic mean-only sensitivity &2&2&2&0\\\bottomrule
\end{tabular}
\end{table}
All references resolved. Every certified count/orientation and every reported invalidity agreed with its reference. Primary yield on admissible inputs is $169/174=97.13\%$; the five remaining cases abstain conservatively near horizon/fold boundaries. Tight primary yield is $136/174$; hints-off retains $169/174$. Grid4097 disagrees on 11/180 primary synthetic requests, and on none of the 90 extension or 30 printed-point contracts. No general accuracy or speed superiority follows.

Fourteen of 28 perturbation pairs satisfy checked sufficient margins; all granted shape claims agree with reference. The other 14 receive no stability conclusion, although ten still preserve shape. Three strict-equality margin probes correctly decline a guarantee. Twenty-seven meaningful certificate alterations are rejected and six benign content transformations accepted; three JSON-order no-ops are identity controls only. This is not a full-suite mutant-detection rate. Primary supervisor wall median is 53.29 ms; the corresponding printed-point median is 52.75 ms. Counters overlap and timings are descriptive, with full costs in Supplement S9--S10.

The printed-point audit certifies all 16 proposed new-source points from three publications, separately from its 14 secondary points. Thirty producer-free CLI rechecks succeed. The 30 contracts comprise 12 root-free and 18 one-simple-root cases; six primary rational-logistic entries are constant-rate GO embeddings. No new published two-root/endpoint/double-root coverage is claimed. 

Screening produced 73 primary records and four additional ZFR records. Together, these comprised 37 INCLUDE, 20 EXCLUDE and 20 UNRESOLVED records. Of the 37 INCLUDE records, 30 had supported encodings and entered the printed-point campaign; seven were held for unsupported month or CPU-time units. The 20 UNRESOLVED records remained outside that campaign. The screened records include formula-level placeholders and are not independent datasets. Eligibility details are in Supplement S10.1.

Rounding robustness was not assessed in this campaign. Shared histories do not establish dataset independence, and the two mean-only sensitivity cases do not validate NHPP/Cox/SDE laws. Shared histories remain conservatively grouped, clock assumptions remain conditional where sources conflict, and audit horizons do not establish unreported estimation spans. Reviewer-reported fit-metric discrepancies remain provenance notes rather than independently reproduced findings. Supplement S12--S13 provides the case identifiers, individual source reservations, review conditions and complete traceability.

\hypertarget{discussion-related-work-and-threats}{%
\section{Discussion, related work and
threats}\label{discussion-related-work-and-threats}}

\hypertarget{what-the-audit-adds}{%
\subsection{What the audit adds}\label{what-the-audit-adds}}

The audit links each claim to the contract obligations and evidence that support it. For applications, this provides checks over the whole declared domain and horizon, explicit failure outcomes, and traceability to sources, code and results.

The certificate route allows a supported contract to be rechecked from mathematical witnesses without the producer or an expected-count label. Yield measures how often bounded discovery obtains those witnesses. The checker's obligations determine what an accepted certificate supports. Primary yield falls from $169/174$ under the default budget to $136/174$ under the tight budget (Section 6.5). This change concerns evidence discovery: failure to find a certificate does not establish model invalidity. Producer-free rechecking succeeds for 30 printed-point certificates. Together with Box 1, these results show how verification can be replayed within the stated parser and backend trust boundary (Supplement S7 and S9--S11).

The final baseline comparison supports additional record rejection within the
mathematical-disagreement and malformed subsets. E2-B demonstrates
preservation under specified representation and clock transformations.
B2 also exposes the interface's distinction between numerical
equivalence and literal context binding. The fixed-front-end rule
ablation further separates record-rejection contributions across rule
groups; zero exclusive rejection gain for E/R in the
mathematical-disagreement category is retained. Its interpretation
follows Section 5.2. Effects on analyst decision quality and analyst effort require separate evaluation.

The code can correctly abstain while a fitted classification remains
analytically decidable at greater precision. This conservative record
outcome is useful only when its reason and component scope remain
visible. Likewise, accepting an unresolved L3 record prevents population
confidence from being inferred from numerical success. 

The residual misses reveal several limits of the current record predicates. Most involve parameter-domain, witness, initial-condition, representation or horizon fields. A field name alone does not show whether an accepted change alters the mathematical application, weakens its justification or affects only its encoding. Resolving that distinction requires obligation-specific adjudication. The results measure integrity sensitivity within the tested corpus; they do not establish exhaustive semantic validation. No tuning follows the final comparison. Tuning to this corpus would require new confirmation. Future work will assess obligation-matched diagnosis by linking reference-defined defects to the checks that actually failed.

\hypertarget{why-population-inference-can-remain-unresolved}{%
\subsection{Why population inference can remain
unresolved}\label{why-population-inference-can-remain-unresolved}}

Inflection count can change discontinuously at endpoint crossings and folds. A sup-norm ball around the contrast admits perturbations that need not correspond to correlated changes in the model parameters. A radius set by behavior away from a root may also be inefficient near a low-slope region. The normalization \(\mathcal H/(1+|\mathcal H|)\) has derivative one at zero, so saturation alone does not explain this root inefficiency. Increasing the number of bootstrap draws reduces Monte Carlo noise. It does not establish the missing uniform-coverage or selection premise.

An alternative is to project a separately justified parameter confidence set through the exact shape classification. If \(\Pr(\theta_0\in C(Y))\ge 1-\alpha\), the full classified image of \(C(Y)\) contains the true count with at least that probability. A numerical outer superset preserves this inclusion; a sampled inner image does not. Likelihood cutoffs and multistart fitting alone do not establish the required premises. Supplement S4 retains exploratory high-count, profile-likelihood and Bayesian results. Neither width contraction nor weak Fisher directions alone proves structural nonidentifiability. Supplement Proposition S4.2 establishes structural identification of the reduced rational-logistic coordinates from the full mean curve when $p$, origin and units are known and $k\ne1$. It does not establish identification from finite grouped counts. The negative E3 results apply to the specified procedure and tested design. Calibration of alternative procedures remains open.

\hypertarget{position-relative-to-prior-work}{%
\subsection{Position relative to prior
work}\label{position-relative-to-prior-work}}

Classical/unified/change-point SRGMs \cite{ref1,ref2,ref3,ref6,ref7} construct and fit
failure means. Detection-rate and generalized-inflection work
\cite{ref8,ref9,ref10} supplies the formulas/sign domains used here.
Imperfect-debugging and mean-interaction models \cite{ref12,ref13,ref14,ref17}
distinguish observation/process semantics. Rana et al.~\cite{ref15} use
empirical profiles for selection and prediction; Chen et al.~\cite{ref16}
develop shape-constrained estimation and pointwise intensity
convergence. Our question is the support for a specific post-fit
validity/shape claim on an explicit horizon. Pointwise intensity
convergence does not supply uniform curvature-root inference without
further argument.

Certifying computation \cite{ref23}, assurance cases \cite{ref24} and provenance
\cite{ref11} inform the decision/evidence contract; mutation work
\cite{ref25,ref26} motivates property-specific error testing.
Reproducibility/FAIR principles \cite{ref27,ref28} inform preservation.
Bootstrap/boundary and feature-inference results \cite{ref18,ref19,ref20,ref21,ref22} motivate
separate calibration obligations. Prediction-domain rate risks in
polynomial models \cite{ref30} and paired semantics in fault-reduction
models \cite{ref31} provide context outside the original map. The
contribution is their scoped executable composition and differentiated
evaluation.

Retrospective source checks qualify NHPP-domain statements, reclassify
printed fitted tuples and reproduce a reported finite-MLE failure. They
do not establish that selected authors counted endpoints as inflections,
confused rate and mean or extrapolated every targeted claim. The review
is source-linked motivation without prevalence estimates; details remain
in Supplement S5.

\hypertarget{threats-and-trust-assumptions}{%
\subsection{Threats and trust
assumptions}\label{threats-and-trust-assumptions}}

\textbf{Oracle/implementation dependence.} Independent code still shares
mathematical assumptions and trusted libraries. Cache-reference
annotation adds no new symbolic derivation for absent inputs. Its
adapter matches the earlier 10,440 comparisons but is not formally
verified. Reference/resolved yield is explicitly retained. Formal and runtime assurance retain the boundaries in Section 3.4; the two human source-contract reviews (Supplement S12) do not validate the parser or arithmetic.

\textbf{Mutation-distribution validity.} New mutations use the same 87
fixtures and defined field operations. Semantic labels and provenance
labels have different targets. The final comparison is post-freeze
analysis, not prospective validation on new families. Metadata controls
and E2-B cover only the specified serializations, exact-value encodings
and two clock-covariant transformations; they do not establish
acceptance of every mathematically equivalent reparameterization.
Mathematical consistency does not make the full record benign. Perfect
observed labelled-subset rejection, absent unchanged-control rejection
and cluster intervals cannot establish population defect rates or
general equivalence acceptance.

\textbf{Numerical and generator truth.} Bounded multistart eligibility
is not global optimality; source/bound sensitivities have not received
separate nested calibration. Ideal-fold and binary-rounded truth differ
as documented. Small finite cell sizes and non-simultaneous intervals
limit inference. Interval checks depend on enclosures and supported
encodings, with abstention on unknown comparisons.

\textbf{External validity/provenance.} Three purposively chosen
published tables do not represent all failure processes. Double entry,
original cited-data checks and sharing permissions remain incomplete.
TROPICO's phases are omitted by the smooth fixed models; NHPP increment
independence is assumed. Unsupported rate families, time-varying fault totals, uncertain population parameters and effort-clock problems lie outside the contract. The purposive source review is not exhaustive. Source screening retained unsupported-unit and unresolved records outside the 30-contract execution cohort; remaining case-specific reservations limit interpretation as detailed in Section 6.5 and Supplement S10--S13. Fixed exact point values of $p$ are supported; a population uncertainty region for $p$ is not automatically supplied.

\hypertarget{conclusions}{%
\section{Conclusions}\label{conclusions}}

For RQ1, the audit states the obligations for validity and fitted shape on a finite horizon. Its guarantees cover conditional classification, certificate soundness, stability of separated simple roots and completeness of the specified reference search. A supported contract and certificate can be rechecked for count and orientation without the producer or an expected-count label. The concrete implementation remains subject to the trust boundary in Section 3.4.

For RQ2, the two routes provide distinct evidence. The primary certificate campaign certifies 169/174 admissible inputs. The record audit rejects all 840 reference-labelled mathematical-disagreement records in the frozen corpus and all 234 composite-challenge records. These results measure yield and record rejection within their stated scope. They do not measure whether each rejection diagnoses the reference-defined defect. The audit also retains 424 changed-field records, classified as integrity misses under the stated metric. Their count does not determine how many are harmful semantic errors.

For RQ3, correctly classifying a selected fitted curve does not ensure that the evaluated band/projection procedure is informative or calibrated. False singleton decisions and broad projected sets limit this procedure under the tested fitting policy and calibration design. These findings do not establish failure of profile-likelihood, Bayesian or other inference procedures. The published histories illustrate why fitted features and supported population conclusions must be reported separately.

Further work should adjudicate reference-consistent rejections and retained misses, measure obligation-matched diagnosis, and evaluate stronger external checkers, new cases and alternative inference procedures. Overall false-rejection rates beyond the specified controls remain unmeasured, as do the semantic consequences of retained changes. Within its declared scope, the audit provides a replayable way to check SRGM applications and report supported finite-horizon fitted-shape conclusions with their evidence.

\hypertarget{data-and-software-availability}{%
\section*{Data and software
availability}\label{data-and-software-availability}}

\begingroup\raggedright
The reproducibility package (record version 1.2.1; \mbox{\url{https://doi.org/10.5281/zenodo.23589336}}) contains the updated supplement (version 1.2.1), a README and checksums, together with the computational archive \path{Paper_A_Zenodo_Reproducibility_v1.2.0.zip}. That archive holds the E1--E4 computational records, exact-contract software and schemas, certificate campaigns, formal sources and source-verification records. Version 1.2.1 revises the supplement, including wording clarifications and the structural-identification proof; computational results are unchanged from version 1.2.0 (\mbox{\url{https://doi.org/10.5281/zenodo.23564540}}). The post-freeze composite-mutation challenge and executed checks are supplied in a separate submission evaluation addendum, outside either archived release. The manuscript is maintained separately and is not included in that archive. After extraction, \path{README.md} at the package root gives the entry points; Supplement S6 and S11 provide execution instructions, and S12--S13 locate the source-review records. Source PDFs and reviewer DOCX forms are not redistributed; reviewer CSV assessment records, reconciliation data and content hashes are included. Original software is licensed under MIT, and author-owned documentation and original research outputs under CC BY 4.0. Third-party source-derived materials and reviewer-authored records retain their respective rights; the scope and exclusions are stated in \path{RIGHTS.md}.
\par\endgroup

\end{document}